\documentclass[conference]{IEEEtran}
\IEEEoverridecommandlockouts

\usepackage{cite}
\usepackage{amsmath}
\usepackage{bbold}
\usepackage{algorithmic}
\usepackage{graphicx}
\usepackage{xcolor}
 \usepackage{array,multirow}
 \usepackage{float}
\def\BibTeX{{\rm B\kern-.05em{\sc i\kern-.025em b}\kern-.08em
    T\kern-.1667em\lower.7ex\hbox{E}\kern-.125emX}}
\begin{document}

\title{CellDefectNet: A Machine-designed Attention Condenser Network for Electroluminescence-based Photovoltaic Cell Defect Inspection
}

\author{Carol Xu$^{2,*}$, Mahmoud Famouri$^{2,*}$, Gautam Bathla$^{2,*}$, Saeejith Nair$^{1}$ \\Mohammad Javad Shafiee$^{1,2*}$, and Alexander Wong$^{1,2*}$ \\
$^1$University of Waterloo, Waterloo, Ontario, Canada\\
$^2$DarwinAI,  Waterloo, Ontario, Canada \\
$^*$Equal Contribution
}

\maketitle

\begin{abstract}
Photovoltaic cells are electronic devices that convert light energy to electricity, forming the backbone of solar energy harvesting systems.  An essential step in the manufacturing process for photovoltaic cells is visual quality inspection using electroluminescence imaging to identify defects such as cracks, finger interruptions, and broken cells.  A big challenge faced by industry in photovoltaic cell visual inspection is the fact that it is currently done manually by human inspectors, which is extremely time consuming, laborious, and prone to human error.  While deep learning approaches holds great potential to automating this inspection, the hardware resource-constrained manufacturing scenario makes it challenging for deploying complex deep neural network architectures.  In this work, we introduce CellDefectNet, a highly efficient attention condenser network designed via machine-driven design exploration specifically for electroluminesence-based photovoltaic cell defect detection on the edge.  We demonstrate the efficacy of CellDefectNet on a benchmark dataset comprising of a diversity of photovoltaic cells captured using electroluminescence imagery, achieving an accuracy of $\sim$86.3\% while possessing just 410K parameters ($\sim$13$\times$ lower than EfficientNet-B0, respectively) and $\sim$115M FLOPs ($\sim$12$\times$ lower than EfficientNet-B0) and $\sim$13$\times$ faster on an ARM Cortex A-72 embedded processor when compared to EfficientNet-B0.
\end{abstract}


\section{Introduction}

\label{sec:intro}

Photovoltaic cells are electronic devices that convert light energy to electricity, forming the backbone of solar energy harvesting systems.  The presence of defects such as interconnect damage, cracks, finger interruptions, material defects, and degraded components can lead to poor power conversion efficiencies and even non-functional cells.  As such, an essential step in the manufacturing process is quality inspection to identify such defects.

\begin{figure}
    \centering
    \includegraphics[height=8cm]{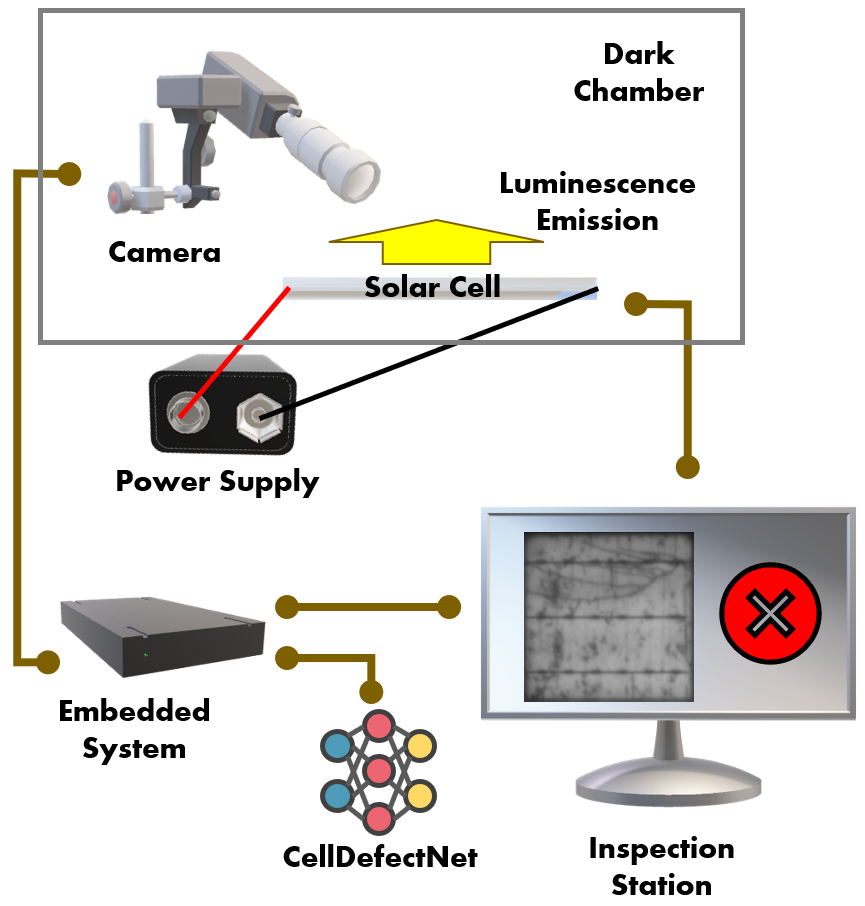}
    \caption{Workflow for computer-assisted electroluminescence-based inspection of photovoltaic cells using CellDefectNet.  A current is fed into the photovoltaic cell in a dark chamber and the radiative recombination of carriers results in luminescence emission.  This luminescence emission is captured using a camera unit, and the captured electroluminescence image is then processed by the proposed CellDefectNet on an embedded system to predict whether the photovoltaic cell is defective or not.  This information can then also be displayed to the human inspector at the inspection station.
    }
    \label{fig:mainflow}
\end{figure}
\begin{figure*}[h]
    \centering
    \includegraphics[width=17cm]{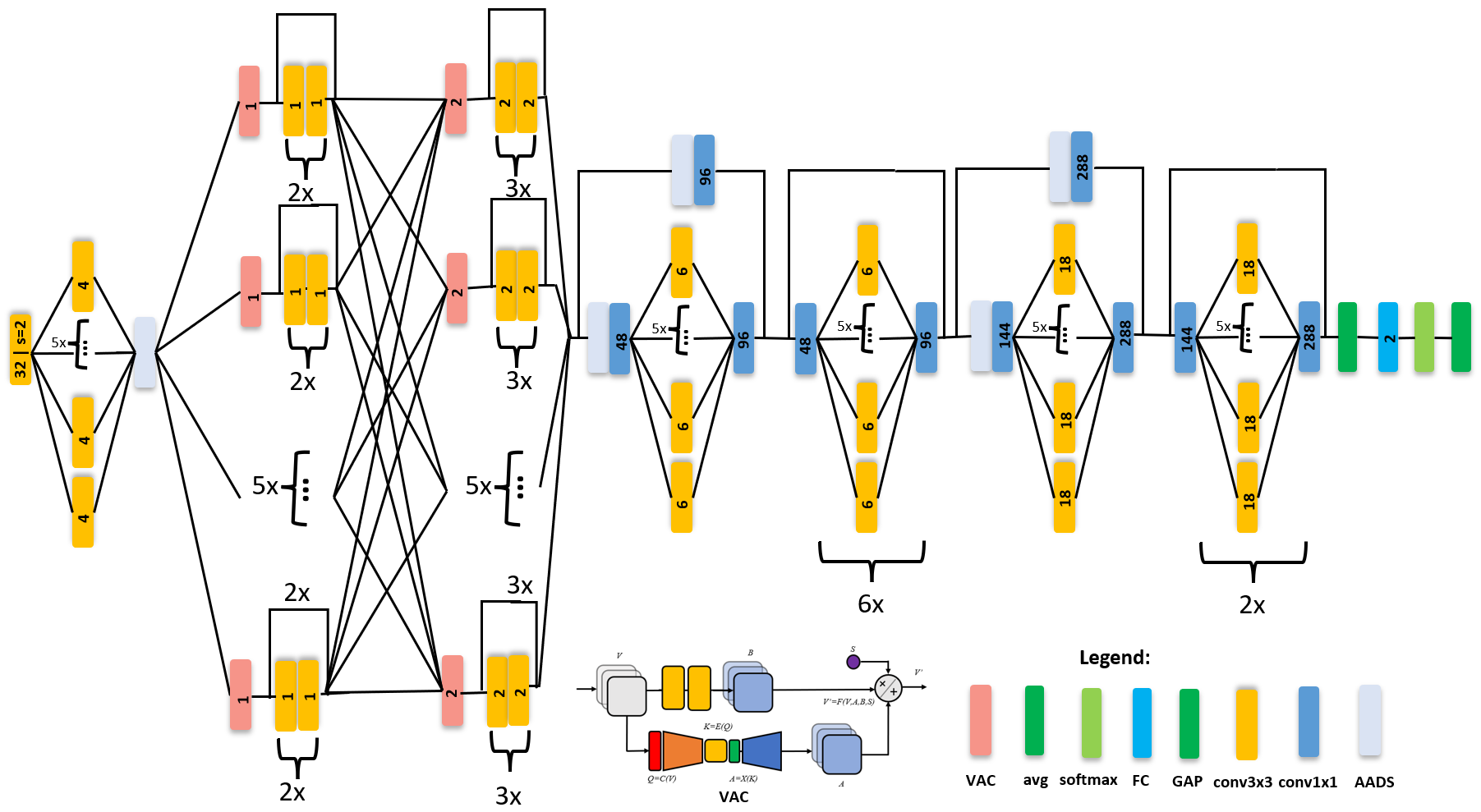}
    \caption{CellDefectNet architectural design.
    The proposed machine-designed attention condenser network architecture design is created via  machine-driven design exploration, and exhibits a heterogeneous columnar design with a large number of columns, early-stage visual attention condensers (VAC) for selective attention, and anti-aliased downsampling (AADS) throughout the network to improve robustness while providing high modeling accuracy.  GAP denotes global average pooling.
    }
    \label{fig:network}
\end{figure*}

Deep learning techniques~\cite{robinson2020learning,he2016deep,reiss2021panda,bergman2020classification} have been showing promising results in different fields and applications. The performance records are being broken by new extensions and improvements everyday. These encouraging results  have motivated the development of new high performing deep neural networks and using these techniques to offer new futuristic solutions such as computer vision applications~\cite{wong2019yolo,redmon2016you,he2017mask}, speech recognition~\cite{amodei2016deep,nassif2019speech} or even for natural language processing tasks~\cite{wolf2020transformers,brown2020language}. This even motivates researchers in the  field of manufacturing to improve the automation including by developing  different manufacturing tasks including the inspection systems~\cite{li2021end,shafiee2021tinydefectnet,9306797}. However, the development in this this area is still in its infancy because several constraints and limitations need to be taken into account for these types of systems including, i) high efficiency and very fast run-time requirements, ii) high accuracy and robustness of the underlying machine learning model, and iii) the limitation on the number of available training data samples of different defected objects. Building deep neural networks satisfying the aforementioned constraints is time-consuming and usually impossible for non-expert users and as such using these systems is still cumbersome.

An effective method for photovoltaic cell quality inspection during manufacturing is the use of electroluminescence imaging, where a current is fed into the photovoltaic cell and the radiative recombination of carriers results in luminescence emission.  While this technique provides greater contrast and clarity of the defects present in a photovoltaic cell to enable improved visual quality inspection during manufacturing, such a process is still conducted manually by a highly trained operator and as such is highly laborious, very time-consuming, and prone to human error.  While deep learning approaches holds great potential to automating this inspection~\cite{deitsch2019automatic}, the hardware resource-constrained manufacturing scenario makes it challenging for deploying complex deep neural network architectures.

Here, we introduce CellDefectNet, a highly efficient attention condenser network designed via machine-driven design exploration specifically for electroluminesence-based photovoltaic cell defect detection on the edge. The inspection workflow using CellDefectNet is shown in Figure~\ref{fig:mainflow}.

\section{Methodology}
The concept of Generative Synthesis~\cite{wong2018ferminets} is utilized to identify the macro- and micro-architecture designs of the proposed CellDefectNet in an automatic approach. The Generative Synthesis process  formulates  the design exploration by a constrained optimization technique. The optimal network architecture is identified  by  an optimal generator $G^\star(\cdot)$. The role of generator $G^\star(\cdot)$ is to generate network architectures $\{\mathcal{N}_s|s \in  S\}$ maximizing a universal performance function $U$ \cite{wong2019netscore} and it is found through the constrained optimization process. The optimization process of finding the generator $G^\star(\cdot)$ is subject to a set of constraints:
\begin{align}
      G^{\star} = \underset{G^{'}} {\max} U \Big(G(s)\Big) \;\;\; \text{s.t.} \;\;\; \mathbb{1}_g(G(s)) = 1 \;\;\; \forall s \in S,
\end{align}
where $S$ is a set of seeds. The operational requirements for the interested network architecture is defined via this set of constraints  formulated via an indicator function $\mathbb{1}_g(\cdot)$.
The generative synthesis process is an iterative approach. At each iteration, the generator  $\bar{G}(\cdot)$ is assessed by an inquisitor $I$ and by generating a set of architectures $\mathcal{N}_s$. The generator at each iteration is evaluated based on the universal performance function $U$ by an indirect evaluation process.

 We take into account several computational and best-practice constraints which are formulated via the indicator function $\mathbb{1}_g(\cdot)$: i) the macroarchitecture design uses several parallel columns to significantly reduce the architectural and computational complexity with much greater disentanglement of learned features; ii) to reduce the considerable information loss caused by the pointwise strided convolutions used in residual networks~\cite{he2016deep} and  RegNet architecture~\cite{radosavovic2020designing} here we restricted its use from the optimization; iii) antialiasing downsampling (AADS)~\cite{zhang2019making} modules are to be used in the early network stage to improve network stability and robustness; iv) FLOPs within 20\% of 100M FLOPs for edge compute scenarios. In the machine driven design exploration process, attention condensers (VAC)~\cite{wong2020tinyspeech, wong2020attendnets} are used as an highly efficient self-attention module to learn and produce condensed embedding characterizing the joint local and cross-channel activation relationships. However, the machine-driven design exploration process automatically determines the best way to satisfy the defined constraints in designing the CellDefectNet architecture.

\subsection{Network Architecture Design}
\vspace{-0.05in}
Figure~\ref{fig:network} demonstrates the CellDefectNet network architecture designed via machine driven exploration. The proposed architecture takes advantage of heterogeneous columnar design with a large number of columns, AADS, and early-stage visual attention condensers (VACs) to improve the robustness while providing the proper modeling accuracy.  A number of key observations can be made about the generated CellDefectNet architecture:
 \begin{enumerate}
     \item {\bf Early-stage self-attention:} the VACs are leveraged heavily within the initial modules used in the network architecture.  VAC was first introduced by Wong \textit{et al.}~\cite{wong2020attendnets} for image classification. The VACs can help to  better model activation relationships and improves selective attention. However VACs adds a very  low complexity to the network compared to other self-attention mechanism. As such, it makes it very attractive for manufacturing use case with real-time processing constraint.
     Utilizing visual attention condensers in these early stages of the network helps to perform selective focus on important low-to-medium level visual indicators. At the same time it improves  the representational efficiency at the very early stages.
     \item \textbf{Heterogeneous columnar design:} the resulted network architecture by the machine-driven exploration uses a heterogeneous combination of columnar design patterns.  The columnar design  enforces a great balance between representational power and disentanglement among the representational feature while maintains the network efficiency. This is amplified by using  more independent columns in early stage of the network with higher interactions in the later stages. This efficient design is well-suited for manufacturing edge computing scenarios as it reduces the computational complexity in the middle of the network while mandating the representational efficacy of the embedding space.
     \item \textbf{Anti-aliased design:} the generated CellDefectNet architecture by the machine-driven exploration technique takes advantage  of anti-aliased downsampling (AADS) operations across the architecture as well. The AADS modules helps the network architecture to improve robustness and stability. Using AADS modules can across the network serves the purpose of conventional pooling operation while account for greater representational efficacy.
 \end{enumerate}

\begin{figure*}
    \includegraphics[width=17cm]{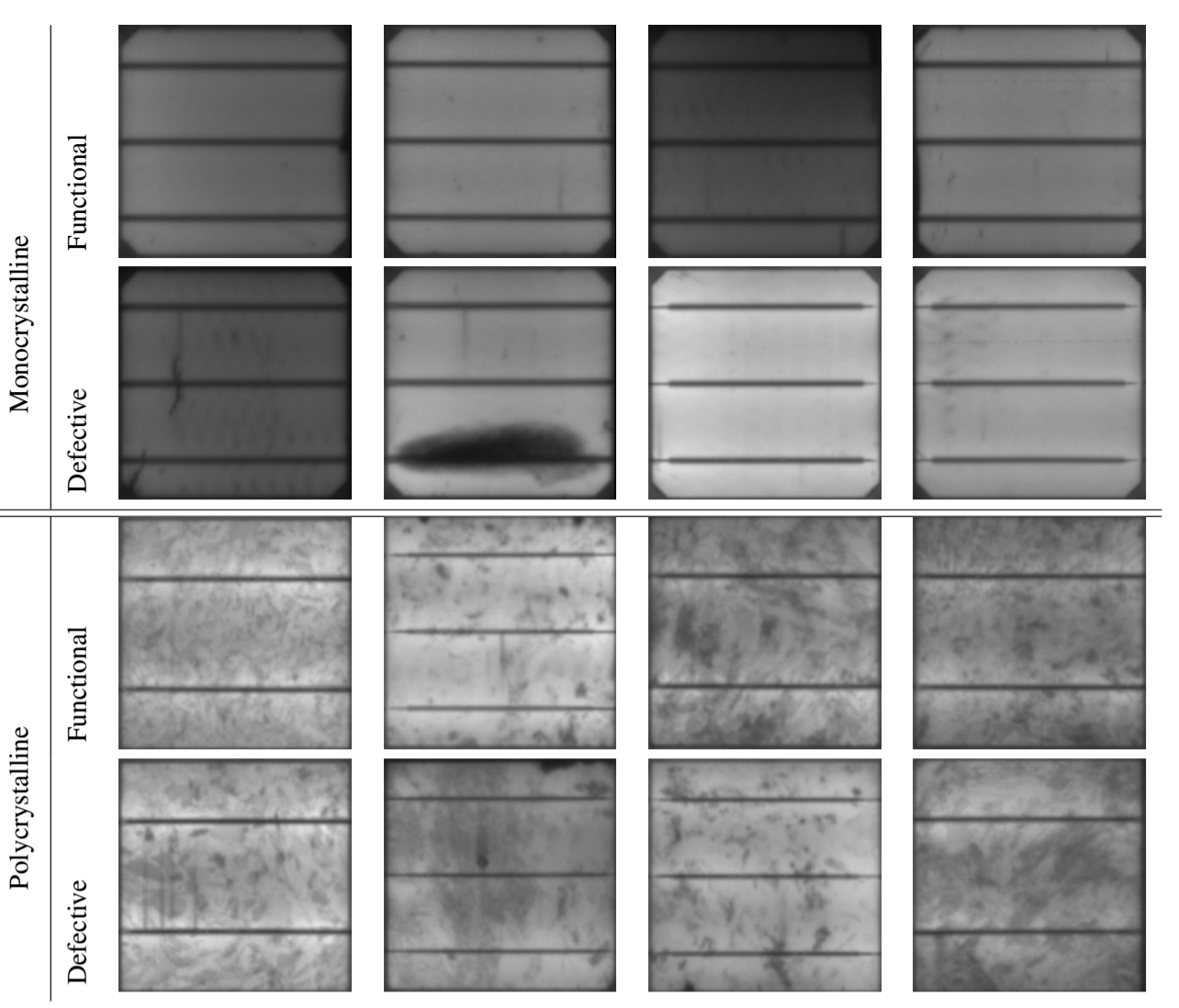}
 \caption{Image samples of the benchmark dataset of photovoltaic cells captured using electroluminescence imagery. The images are from two different types of solar cells including, Monocrystalline and Ploycrystalline. As seen, it is very difficult to distinguish between the functional and defective samples from both types of solar cells.}
 \label{fig:exam}
\end{figure*}

\section{Results \& Discussion}

\subsection{Experimental  Setup}
\label{sec:exp}
To explore the efficacy of the proposed CellDefectNet, we evaluated its performance on a benchmark dataset comprising of a diversity of photovoltaic cells captured using electroluminescence imagery~\cite{deitsch2019automatic}.  The benchmark dataset comprises of 2624 images captured of monocrystalline and polycrystalline photovoltaic cells with 1100 defective samples and 1514 non-defective samples, with a training/test split of 75\%/25\% as described in~\cite{9306797}.  The resolution of the images are 300 $\times$ 300.

Figure~\ref{fig:exam} demonstrates samples from the benchmark dataset and different possible defects. As seen, it is very difficult to distinguish between the defect-free (functional) and defective samples without having expert knowledge to identify them.

\subsection{Competing Methods}
In addition to evaluating the performance of CellDefectNet, we also tested on the VGG-19~\cite{Simonyan2015VGG19} network architecture on the same benchmark for reference purposes given that it was used as a backbone for photovoltaic defect detection in~\cite{deitsch2019automatic}, and we also evaluated two state-of-the-art efficient architectures: EfficientNet-B0~\cite{tan2020efficientnet} and MnasNet~\cite{tan2019MnasNet}.  As described in~\cite{deitsch2019automatic}, the network architectures were first trained using Adam optimization on only the fully-connected layers for 100 epochs with a learning rate of $1.0\times10^{-3}$, then trained with stochastic gradient descent optimization on the entire network for 100 epochs with a learning rate of $5.0\times10^{-4}$ and batch size of 16.
\begin{center}
    \begin{table*}
    \center
        \caption{ Quantitative results of the proposed CellDefectNet architecture compared to other tested architectures.
        It can be seen that the proposed CellDefectNet architecture is as much as $\sim$341.46$\times$ smaller in terms of parameters and as much as $\sim$300$\times$ lower computational complexity in terms the number of FLOPs. Furthermore, CellDefectNet is as much as 31.39$\times$ faster on an ARM Cortex A-72 64-bit 1.5GHz embedded processor.
        }

    \begin{tabular}{l || c c c c }
        Model &  Test Acc (\%) &Param (M) & FLOPs (M) & Run-time (s)\\
        \hline \hline
         VGG-19 & 85.06 & 140 & 34570 & 10.893 \\
        \hline
         EfficientNet-B0 & 85.37 & 5.3 & 1397 & 4.479\\
        \hline
         MnasNet & 83.69 & 3.9 & 1074 & 3.746\\
        \hline
        CellDefectNet & \textbf{86.28} & \textbf{0.41} & \textbf{115} & \textbf{0.347}
    \end{tabular}
    \label{tab:res}
    \end{table*}
\end{center}

\subsection{Performance Metrics and Results}
The competing models are evaluated based on  their quantitative performance, architectural complexity, computational complexity, and inference speed. For quantitative performance, we compute the test accuracy as the key metric.  For architectural complexity, we assess the number of parameters in the neural network architecture as the key metric.  For computational complexity, we measure the number of floating point operations (FLOPs) as the key metric.  Finally, for inference speed, we measure the run-time latency on an ARM Cortex A-72 64-bit 1.5GHz processor as the key metric. The comparative results for the proposed CellDefectNet and several state-of-the-art efficient architectures are illustrated  in Table~\ref{tab:res}.

\subsection{Architectural Complexity}
First, from an architectural complexity perspective, \mbox{CellDefectNet} consists of just $\sim$410K parameters, which is significantly smaller than all the competing state-of-the-art efficient architectures.  More specifically, CellDefectNet is $\sim$341.46$\times$ smaller compared to the VGG-19 architecture while achieving 1.2\% higher accuracy.
CellDefectNet is $\sim$13$\times$ smaller compared to the highly efficient EfficientNet-B0 (the most accurate architecture outside of CellDefectNet) while achieving $\sim$0.9\% higher accuracy.  Furthermore, CellDefectNet is $\sim$9.5$\times$ smaller compared to the highly efficient MnasNet (the most efficient architecture outside of CellDefectNet) while achieving $\sim$2.6\% higher accuracy.

\subsection{Computational Complexity}
Second, in terms of computational complexity, CellDefectNet requires only $\sim$115M FLOPs, which is significantly lower than VGG-19 as well as the tested state-of-the-art efficient architectures.  More specifically, CellDefectNet requires $\sim$300$\times$ fewer FLOPs compared to the VGG-19 architecture,  $\sim$12.14$\times$ fewer FLOPs compared to EfficientNet-B0, and $\sim$9.34$\times$ fewer FLOPs compared to MnasNet.

\subsection{Accuracy}
Third, from an accuracy perspective, CellDefectNet achieved the highest accuracy amongst  the tested architectures at a test accuracy of 86.28\%.  These results illustrate the strong balance achieved by CellDefectNet in terms of accuracy, architectural complexity, and computational complexity, making it very well-suited for high-performance Photovoltaic cells defect detection in resource-constrained manufacturing environments.

\subsection{Embedded inference speed} We further explore real-world operational efficiency of the proposed CellDefectNet architecture in embedded scenarios by evaluating its run-time latency (at a batch size of 1) on an ARM Cortex A-72 64-bit 1.5GHz processor in comparison with the other tested architectures in this study.  It can be seen from Table~\ref{tab:res} that the proposed CellDefectNet architecture is able to achieve a runtime latency of 0.347~s per sample, which is significantly lower than that of VGG-19 along with the tested state-of-the-art efficient deep neural network architectures explored in this study.  More specifically, the proposed CellDefectNet is 31.39$\times$ faster when compared to the VGG-19 architecture, $\sim$13$\times$ faster when compared to the EfficientNet-B0 architecture, and $\sim$10.8$\times$ faster when compared to the MnasNet architecture (the fastest architecture running on the ARM embedded processor outside of CellDefectNet).  The significant speed advantages of CellDefectNet make it very well-suited for use on embedded edge compute devices for high-throughput manufacturing scenarios.  Furthermore, the significantly lower architectural and computational complexity as well as higher accuracy achieved by the proposed CellDefectNet network architecture illustrates the effectiveness of leveraging a machine-driven design exploration strategy with both computational as well as ``best-practices'' constraints in the creation of highly tailored deep neural network architectures that are designed and customized in an automatic fashion for a specific industry task and scenario at hand and on the edge.

\section{Conclusions}

Here, we took advantage a machine-driven design exploration with computational and ``best-practices`` to build a highly compact deep neural network architectures for the task of photovoltaic cell defect detection.  The resulting network architecture, so-called CellDefectNet, contains a unique self-attention network architecture with heavily usage of heterogeneous columnar macro-architecture design, antialiasing properties, and highly tailored microarchitecture design to strongly balance between accuracy, robustness, and efficiency for the real-world manufacturing use cases.  Experimental results shows  that the proposed CellDefectNet is able to achieve a detection accuracy of $\sim$86.2\% on the photovoltaic cells captured using electroluminescence imagery benchmark with highly efficient architectural and very lower computational complexity when compared to state-of-the-art efficient deep neural network architectures.  Furthermore, the run-time experiments demonstrates that the proposed CellDefectNet achieves significantly faster inference speed on an embedded ARM processor with 31.39$\times$ speed-up compared to the state-of-the-art model for this purpose, a very well-suited machine learning model for photovoltaic cell defect detection in high-throughput, resource-constrained manufacturing scenarios. The future works aims to explore and leveraging of this machine-driven design exploration strategy to build even more efficient architecture for this problem and also produce highly efficient yet high-performing deep neural network architectures for other critical manufacturing applications with using different sensing modalities such as acoustic sensors for predictive maintenance.

{\small
\bibliographystyle{ieee_fullname}
\bibliography{egbib}

\begin{thebibliography}{10}\itemsep=-1pt

\bibitem{amodei2016deep}
Dario Amodei, Sundaram Ananthanarayanan, Rishita Anubhai, Jingliang Bai, Eric
  Battenberg, Carl Case, Jared Casper, Bryan Catanzaro, Qiang Cheng, Guoliang
  Chen, et~al.
\newblock Deep speech 2: End-to-end speech recognition in english and mandarin.
\newblock In {\em International conference on machine learning}, pages
  173--182. PMLR, 2016.

\bibitem{bergman2020classification}
Liron Bergman and Yedid Hoshen.
\newblock Classification-based anomaly detection for general data.
\newblock {\em arXiv preprint arXiv:2005.02359}, 2020.

\bibitem{brown2020language}
Tom Brown, Benjamin Mann, Nick Ryder, Melanie Subbiah, Jared~D Kaplan, Prafulla
  Dhariwal, Arvind Neelakantan, Pranav Shyam, Girish Sastry, Amanda Askell,
  et~al.
\newblock Language models are few-shot learners.
\newblock {\em Advances in neural information processing systems},
  33:1877--1901, 2020.

\bibitem{deitsch2019automatic}
Sergiu Deitsch, Vincent Christlein, Stephan Berger, Claudia Buerhop-Lutz,
  Andreas Maier, Florian Gallwitz, and Christian Riess.
\newblock Automatic classification of defective photovoltaic module cells in
  electroluminescence images.
\newblock {\em Solar Energy}, 185:455--468, 2019.

\bibitem{he2017mask}
Kaiming He, Georgia Gkioxari, Piotr Doll{\'a}r, and Ross Girshick.
\newblock Mask r-cnn.
\newblock In {\em Proceedings of the IEEE international conference on computer
  vision}, pages 2961--2969, 2017.

\bibitem{he2016deep}
Kaiming He, Xiangyu Zhang, Shaoqing Ren, and Jian Sun.
\newblock Deep residual learning for image recognition.
\newblock In {\em Proceedings of the IEEE conference on computer vision and
  pattern recognition}, pages 770--778, 2016.

\bibitem{li2021end}
Yue Li and Junfeng Li.
\newblock An end-to-end defect detection method for mobile phone light guide
  plate via multitask learning.
\newblock {\em IEEE Transactions on Instrumentation and Measurement}, 70:1--13,
  2021.

\bibitem{9306797}
Zhaopan Li, Junfengs Li, and Wenzhan Dai.
\newblock A two-stage multiscale residual attention network for light guide
  plate defect detection.
\newblock {\em IEEE Access}, 9:2780--2792, 2021.

\bibitem{nassif2019speech}
Ali~Bou Nassif, Ismail Shahin, Imtinan Attili, Mohammad Azzeh, and Khaled
  Shaalan.
\newblock Speech recognition using deep neural networks: A systematic review.
\newblock {\em IEEE access}, 7:19143--19165, 2019.

\bibitem{radosavovic2020designing}
Ilija Radosavovic, Raj~Prateek Kosaraju, Ross Girshick, Kaiming He, and Piotr
  Doll{\'a}r.
\newblock Designing network design spaces.
\newblock In {\em Proceedings of the IEEE/CVF Conference on Computer Vision and
  Pattern Recognition}, pages 10428--10436, 2020.

\bibitem{redmon2016you}
Joseph Redmon, Santosh Divvala, Ross Girshick, and Ali Farhadi.
\newblock You only look once: Unified, real-time object detection.
\newblock In {\em Proceedings of the IEEE conference on computer vision and
  pattern recognition}, pages 779--788, 2016.

\bibitem{reiss2021panda}
Tal Reiss, Niv Cohen, Liron Bergman, and Yedid Hoshen.
\newblock Panda: Adapting pretrained features for anomaly detection and
  segmentation.
\newblock In {\em Proceedings of the IEEE/CVF Conference on Computer Vision and
  Pattern Recognition}, pages 2806--2814, 2021.

\bibitem{robinson2020learning}
Andreas Robinson, Felix~Jaremo Lawin, Martin Danelljan, Fahad~Shahbaz Khan, and
  Michael Felsberg.
\newblock Learning fast and robust target models for video object segmentation.
\newblock In {\em Proceedings of the IEEE/CVF Conference on Computer Vision and
  Pattern Recognition}, pages 7406--7415, 2020.

\bibitem{shafiee2021tinydefectnet}
Mohammad~Javad Shafiee, Mahmoud Famouri, Gautam Bathla, Francis Li, and
  Alexander Wong.
\newblock Tinydefectnet: Highly compact deep neural network architecture for
  high-throughput manufacturing visual quality inspection.
\newblock {\em arXiv preprint arXiv:2111.14319}, 2021.

\bibitem{Simonyan2015VGG19}
Karen Simonyan and Andrew Zisserman.
\newblock Very deep convolutional networks for large-scale image recognition.
\newblock {\em CoRR}, abs/1409.1556, 2015.

\bibitem{tan2019MnasNet}
Mingxing Tan, Bo Chen, Ruoming Pang, Vijay Vasudevan, and Quoc~V. Le.
\newblock Mnasnet: Platform-aware neural architecture search for mobile.
\newblock {\em CoRR}, abs/1807.11626, 2018.

\bibitem{tan2020efficientnet}
Mingxing Tan and Quoc~V. Le.
\newblock Efficientnet: Rethinking model scaling for convolutional neural
  networks.
\newblock {\em CoRR}, abs/1905.11946, 2019.

\bibitem{wolf2020transformers}
Thomas Wolf, Lysandre Debut, Victor Sanh, Julien Chaumond, Clement Delangue,
  Anthony Moi, Pierric Cistac, Tim Rault, R{\'e}mi Louf, Morgan Funtowicz,
  et~al.
\newblock Transformers: State-of-the-art natural language processing.
\newblock In {\em Proceedings of the 2020 conference on empirical methods in
  natural language processing: system demonstrations}, pages 38--45, 2020.

\bibitem{wong2019netscore}
Alexander Wong.
\newblock Netscore: towards universal metrics for large-scale performance
  analysis of deep neural networks for practical on-device edge usage.
\newblock In {\em International Conference on Image Analysis and Recognition},
  pages 15--26. Springer, 2019.

\bibitem{wong2020tinyspeech}
Alexander Wong, Mahmoud Famouri, Maya Pavlova, and Siddharth Surana.
\newblock Tinyspeech: Attention condensers for deep speech recognition neural
  networks on edge devices.
\newblock {\em arXiv preprint arXiv:2008.04245}, 2020.

\bibitem{wong2020attendnets}
Alexander Wong, Mahmoud Famouri, and Mohammad~Javad Shafiee.
\newblock Attendnets: tiny deep image recognition neural networks for the edge
  via visual attention condensers.
\newblock {\em arXiv preprint arXiv:2009.14385}, 2020.

\bibitem{wong2019yolo}
Alexander Wong, Mahmoud Famuori, Mohammad~Javad Shafiee, Francis Li, Brendan
  Chwyl, and Jonathan Chung.
\newblock Yolo nano: a highly compact you only look once convolutional neural
  network for object detection.
\newblock In {\em 2019 Fifth Workshop on Energy Efficient Machine Learning and
  Cognitive Computing-NeurIPS Edition (EMC2-NIPS)}, pages 22--25. IEEE, 2019.

\bibitem{wong2018ferminets}
Alexander Wong, Mohammad~Javad Shafiee, Brendan Chwyl, and Francis Li.
\newblock Ferminets: Learning generative machines to generate efficient neural
  networks via generative synthesis.
\newblock {\em arXiv preprint arXiv:1809.05989}, 2018.

\bibitem{zhang2019making}
Richard Zhang.
\newblock Making convolutional networks shift-invariant again.
\newblock In {\em International conference on machine learning}, pages
  7324--7334. PMLR, 2019.

\end{thebibliography}
}
\end{document}